\documentclass[preprint,10pt,number]{elsarticle}
\usepackage[utf8]{inputenc}
\usepackage{amssymb}
\usepackage{amsthm}
\usepackage{amsmath}
\usepackage{adjustbox}
\usepackage[section]{placeins}
\usepackage{color}
\usepackage{wrapfig}
\usepackage{etoolbox,refcount}
\usepackage{multicol}
\usepackage{geometry}
\usepackage{grffile}
\usepackage{setspace}
\usepackage{sectsty}
\sectionfont{\fontsize{15}{20}\selectfont}
\subsectionfont{\fontsize{13}{15}\selectfont}
\subsubsectionfont{\fontsize{11}{15}\selectfont}
\usepackage{graphicx}
\usepackage{caption}
\usepackage{subcaption}
\usepackage{xcolor}
\usepackage{datetime}
\usepackage{textcomp}
\usepackage{fontspec}
\newfontfamily\corsiva{Monotype-Corsiva-Regular.ttf}

\begin{document}
\begin{frontmatter}

\title{\Large Towards Engineering Material Neural Networks}

\author{Charles de Kergariou$^a$*, Hortense le Ferrand$^b$, Ali Momeni$^c$, Romain Fleury$^c$, Kunal Masania$^d$, Adam Perriman$^e$$^f$, Fabrizio Scarpa$^a$}
\address{$^a$ Bristol Composites Institute, School of Civil, Aerospace and Mechanical Engineering, University of Bristol, University Walk, Bristol BS8 1TR, UK

$^b$ School of Mechanical and Aerospace Engineering, Nanyang Technological University, Singapore 639798

$^c$ Laboratory of Wave Engineering, School of Electrical Engineering, Ècole Polytechnique Fédérale de Lausanne (EPFL), Lausanne, Switzerland

$^d$ Shaping Matter Lab, Faculty of Aerospace Engineering, Delft University of Technology, Kluyverweg 1, Delft 2629 HS, Netherlands

$^e$ Research School of Chemistry and John Curtin School of Medical Research, Australian National University, Canberra ACT2601, Australia

$^f$ School of Cellular and Molecular Medicine, University of Bristol, University Walk, Bristol BS8 1TD, United Kingdom

*Corresponding author.
Email address: charles.dekergariou@bristol.ac.uk}

\begin{abstract}
Structures that capture functionality in the form of animate or intelligent machines have the potential to transform modern engineering applications. Animation and embedded intelligence are typically realised by integrating advanced capabilities such as reversibility, adaptive responses and learning directly into the materials themselves. Currently, the majority of adaptive material systems rely on predefined adaptive designs combined with in-service, electronics-based computing to dynamically modify the structural behaviour. However, structural configurations with interconnected adaptable nodes are able to approximate continuous functions, providing new possibilities and opportunities than classical metamaterials and computational materials. We discuss here the potential to design load-bearing engineering materials with trainable physical parameters and neural network-inspired morphologies, embedding intelligence directly into their structure a concept we define as Engineering Material Neural Networks (EMNNs) as a subcategory of Physical Neural Networks.
In this perspective we first establish the foundational concept of EMNNs; we then detail the mechanical and multifunctional properties required for such structural configurations. Finally, we evaluate existing and emerging engineering materials that hold promise for enabling this innovative approach. Key material candidates for realising EMNNs include composites, architected, biological and engineering living materials. We also outline future directions in materials science and structural engineering for developing EMNNs.

\end{abstract}

\end{frontmatter}

\noindent\textbf{Keywords:} Engineering Material Neural Networks; Physical Neural Network; Engineering Living Materials; Composite Materials; Microstructured Materials

\begin{figure}[h]
    \centering
    \includegraphics[width=0.9983\textwidth]{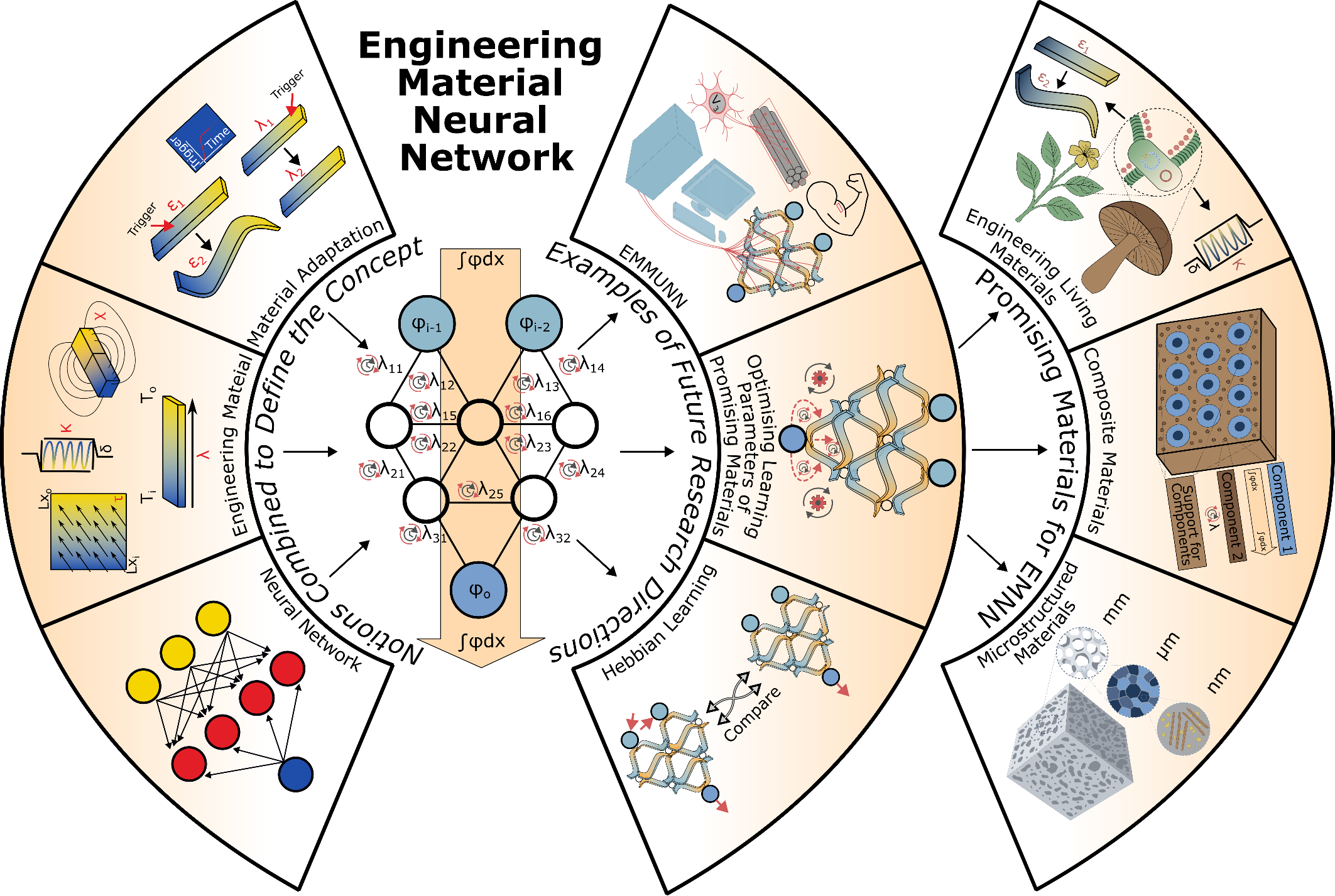}
    \vspace{-3mm}
    \caption*{\textbf{Graphical Abstract} }
    \label{fig: 1}
\end{figure}

\section{Introduction}

Over the past decades, AI has emerged as an indispensable tool across diverse fields including healthcare, software development and design. \cite{Horton2025AcceleratedProject}. New physical computing systems, however, are being developed to limit the use of expensive and non-sustainable graphic processing units (GPU). Since the optical correlator in 1964, structures mimicking neural network behaviour capable of computation have been developed in optics and electronics \cite{VanderLugt1964SignalFiltering, Momeni2025TrainingNetworks}. These structures still rely on the use of external electric power and serve the same computational functions as conventional processing units. The material properties of these neural network-mimicking structures have also been adapted to serve as classifiers \cite{Li2024TrainingBackpropagation}. The intrinsic materials-based approach to design these neural network-like structures enhances the computational energy efficiency, as neuronal connections are not processed by a GPU but they are physically embodied through the material’s stiffness. Physical computing can be traced back to ancient Greece and was largely popularised by Charles Babbage \cite{Yasuda2021MechanicalComputing}. Current improvements in material science and structural design allow to evolve from rigid gear and beam components assemblies to network of materials behaving as logic gates \cite{Yasuda2021MechanicalComputing,Li2024ReprogrammableMemory}. However, structural parameters (e.g. stiffness, density, geometry, conductivity) for the material-based computation are fixed during production and cannot be adapted in service to achieve the desired output or correct errors. Buckled states of beams have been used to create a robust neural network capable of recognising numbers \cite{Mei2024MechanicalNeurons} and performing computing \cite{Mei2023In-memoryComputing}. To accomplish this, the physical system was composed of logic gates as integral components \cite{Mei2021AFunctions}. With this technology, Mei et al. were able to display long/short term memory in the structural configurations they created \cite{Mei2021AFunctions}. However, these adaptive structures are mostly used as a non-digital substitute of computers \cite{Li2024TrainingBackpropagation}; the physical nature of material-based neural network makes operation on information complex.
In electromagnetic wave-based physical neural networks, wave physics provides the computational substrate for mathematical operations, while electronics typically handle modulation, readout, memory or control functions \cite{Wright2022DeepBackpropagation}. When mimicking the brain in structure or function, such systems are called neuromorphic computers \cite{Wright2022DeepBackpropagation}. Similar neuromorphic concepts \cite{Chiappalone2026AdvancingTwins} have been proposed for metamaterials \cite{Sylvestre2021NeuromorphicStructures}. However, the non electronic physical junctions between unit cells (nodes) make it difficult to adapt the computations occurring within the metamaterial. For example, in the neuromorphic system described by Sylvestre \textit{et al.}, the stiffness node junction's computing power is determined by input loads defined during production and cannot be adapted in service \cite{Sylvestre2021NeuromorphicStructures}. Hence, creating adaptable junctions would represent a way to increase the adaptability and computing power of metamaterials. Following the physical control approaches described in \cite{Milana2025PhysicalRobotics}, a structure's behaviour would directly result from the interactions between its materials and the environment, facilitated by the transmission of physical information within the robotic structure. Material-based computing structures could therefore enable real-time, on-the-fly learning of user-specified behaviours.

Here we examine how to advance energy autonomy and matter intelligence in engineering materials, and how this advancement could pave the way for Engineering Materials Neural Networks (EMNNs). The EMNN concept describes a neuromorphic structure that converts physical variables into measurable outputs by approximating a physical function through the architecture presented in Fig. \ref{fig: 1}. An EMNN leverages the intrinsic properties of a material to achieve this transformation. The system's inputs and outputs can be of the same nature or dissimilar, as in the case of transduction functions. EMNNs represent a subcategory of physical neural networks \cite{Momeni2025TrainingNetworks}, consisting of load-bearing architected material structures with trainable physical parameters. Adapting their properties enables the entire structure to exhibit learning behaviour, transforming inputs into task-relevant outputs. Figure \ref{fig: 1} presents the conceptual layout and the different parts of EMNNs.

\begin{figure}[h]
    \centering
    \includegraphics[width=0.683\textwidth]{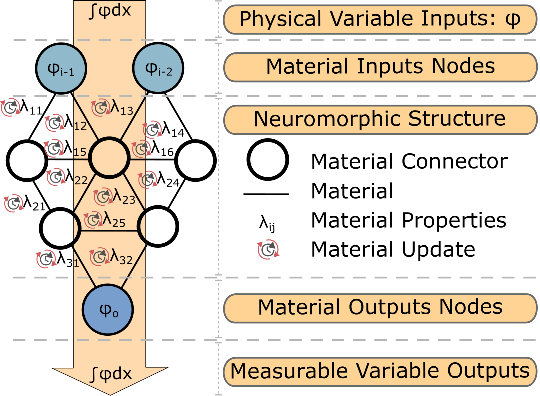}
    \vspace{-3mm}
    \caption{\textbf{Neuromorphic structure and behaviour of an Engineering Material Neural Network.}  The yellow arrow at the back of the neural network represents the physical transformation of variable. Here the variable transformed is $\phi$ and $x$ is the space over which the physical variable is transformed.}
    \label{fig: 1}
\end{figure}

The development of EMNNs will target three main objectives: sustainability, autonomy and optimisation. The engineering material neural network is designed to minimise the energy required for behavioural training and to reduce the time operators spend tuning physical parameters, enabling the structure to autonomously adapt and optimise its performance in unpredictable contexts. NASA plans to use automata for Venus surface exploration \cite{Sauder2017AUTOMATONInvestigator}. EMNNs could offer energy-efficient autonomy and adaptive optimization required for navigating the unpredictable and harsh Venusian environment. Rather than proposing specific solutions, this perspective paper outlines potential pathways toward achieving these neural network-like material systems. The EMNN concept proposed in this perspective also emphasizes the use of load-bearing engineering materials for potential applications in neural network-inspired structures.

We propose that Engineering Materials Neural Networks should exhibit the following three key characteristics:

\begin{itemize}
    \item \textit{Rule \#1}: The structure must feature an assembly of individual interconnected material nodes individually adaptable.
    \item \textit{Rule \#2}: The output of the system must be adaptable according to the demand of the user.
    \item \textit{Rule \#3}: The design of the EMNN should optimise energy efficiency, output accuracy, independence from traditional digital computations and speed of convergence.
\end{itemize}

In EMNNs, learning capabilities should guide hardware design, alongside optimizations for performance, energy efficiency, and speed. Such energy-based design requirements will facilitate scaling up. In this perspective, \textit{autonomous actuation} refers to the ability of a system to obtain the inputs required to operate and to process them through feed the forward inference or pass. \textit{Feedback control} refers to the update of the weight of the physical neural network. \textit{Learning} here is used to designate the optimisation of trainable parameters for a specific task. \textit{Adaptation}, at both the material and structural scales, refers to the change in behaviour or response following an adaptation trigger. In the following subsection we will discuss a structure capable of achieving these rules for engineering materials.

\subsection{Structure}
\begin{figure}[h]
    \centering
    \includegraphics[width=0.9983\textwidth]{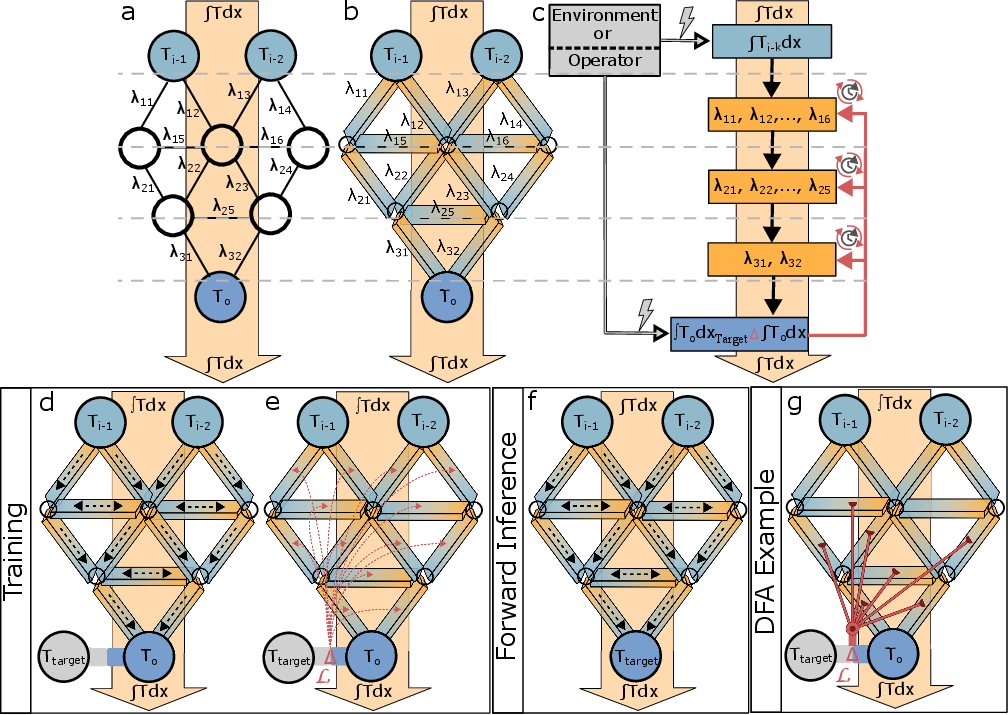}
    \vspace{-3mm}
    \caption{\textbf{Theoretical concept, functions and operations of an EMNN during training and exploitation.} \textbf{a,} Example of Artificial Neural Network (ANN) structure with interconnected nodes. \textbf{b,} Example of Engineering Material Neural Network (equivalent to the ANN) for temperature control. ($\lambda$: Thermal Conductivity; T: Temperature) \textbf{c,} Schematic presenting the information transmission in an EMNN. The clock symbols represent the update of the material parameters. The thunder symbols and the black/silver arrows present the information given by in environment of the operator to the inputs and output of the system. \textbf{d,} Forward Inference pass during the training phase. \textbf{e,} Feedback updating process during the training phase to adapt the material properties. ($\Delta$: Difference between Targetted and Output temperatures as presented in \textbf{c,}; \text{\corsiva L}: Loss calculated with Eq. \ref{eq: 2}) \textbf{f,} Forward inference phase in service. \textbf{g,} Example of training of the structure via DFA.}
    \label{fig: 2}
\end{figure}

Networks of interconnected nodes can approximate continuous functions\cite{Hornik1989MultilayerApproximators}. To adapt to environmental conditions and produce the desired output, a networked structure of interconnected nodes can dynamically approximate the necessary behaviour. Fig. \ref{fig: 2}a presents an example of interconnected nodes of an Artificial Neural Network (ANN) able to adapt the coefficients $\lambda_{ij}$ to obtain the output $T_o$ from the input $T_{i-k}$ in multiple configurations. Fig. \ref{fig: 2}b shows its equivalent EMNN version, in which all parameters become physical to respect the \textit{Rule \#1} previously defined. In this example, the coefficients $\lambda_{ij}$ represent the thermal conductivity, while $T_o$ and $T_{i-k}$ are the output and input temperatures. Fig. \ref{fig: 2}c introduces the training logic of an EMNN. Every arrow represents the transmission of an information, while the clock symbol represents a properties update. The $\Delta$ sign indicates that the system must calculate the difference between target and obtained outputs, and then compute the gradient to update the layer of the EMNN. In additions to \textit{Rules \#1-\#3}, the structure must also fulfil the following functions to behave as an EMMN:

For the Forward Pass (FW) (black arrows in Fig. \ref{fig: 2}c, \ref{fig: 2}d and \ref{fig: 2}f):

\begin{itemize}
    \item \textit{Step FW1}: Encode the information of the input provided by the operator or the environment through physical changes within the EMNN.
    \item \textit{Step FW2}: Propose a physical model for the node grid that evolves through coupled material dynamics under boundary conditions and trainable parameters to generate outputs.
\end{itemize}

From the forward pass, the measurable variable output of the EMNN can be calculated using Eq. \ref{eq: 1}, with \textit{f} being the function approximated via the neural network:

\begin{equation}
    T_o=f(\lambda_{ij};T_i)
    \label{eq: 1}
\end{equation}

For the FeedBack (FB) pass of the network (i.e., the red symbols in Fig. \ref{fig: 2}c and \ref{fig: 2}e):

\begin{itemize}
    \item \textit{Step FB1}: Compare the obtained output against the target. For instance, the loss and the error could be calculated for the forward pass with Eq. \ref{eq: 2} and Eq. \ref{eq: 3}, respectively. One of the core difficulties in creating EMNN lies in the physical obtention of these values. Example of methods to achieve such analysis, such as digital twin, physics aware training, hamiltonian echo backpropagation or eroth-order gradient are presented later in this paper.

\begin{equation}
    \text{\corsiva L} = l (T_o, T_{target})
    \label{eq: 2}
\end{equation}

\begin{equation}
    \nabla_{\lambda} \text{\corsiva L} = \frac{\partial \text{\corsiva L}}{\partial T_o} \cdot \frac{\partial T_o}{\partial \lambda}
    \label{eq: 3}
\end{equation}
    
    \item \textit{Step FB2}: Calculate an updated gradient for the weights.
    \item \textit{Step FB3}: Transfer the information of the gradient to the nodes.
    \item \textit{Step FB4}: Update the weights with the gradient. An example of gradient calculation is given in Eq. \ref{eq: 4}.
\end{itemize}

\begin{equation}
    \lambda \leftarrow \lambda - \mu \nabla_{\lambda} \text{\corsiva L}
    \label{eq: 4}
\end{equation}

A metamaterial is a 3D structure with a response or function due to the collective effect of meta-atom elements that is not possible to achieve conventionally with any individual constituent material. As such, metamaterials can generate functionalities beyond the properties of the classical bulk or continuum materials \cite{Zheng2025MetamaterialRobotics}. Metamaterials also often possess a structure which could be assimilated at repeated pattern of nodes \cite{Dudek2025Shape-morphingMetamaterials,Qian2025AIntelligence,Zhang2025RecentSystems}. Fig. \ref{fig: 2}d, \ref{fig: 2}e and \ref{fig: 2}f illustrate the missing functionalities that metamaterials must acquire to evolve into Engineering Material Neural Networks. Classical metamaterials lack \textit{cognitive} training functions \cite{Jiao2023MechanicalBeyond}. Metamaterials however already feature an exploitation phase, for which forward pass functions are required (FW1, FW2). By incorporating a training phase and adaptation, a metamaterial could respond to changing conditions and discover optimised behavioural patterns even those unforeseen by conventional modelling.

\section{Training Strategies: Autonomous Physical Adaptability}

As described in Fig. \ref{fig: 1}, the main objective of a EMNN is to transform a physical variable into the required output. The real-time, adaptive functionality of EMNNs depends solely on the forward inference pass. The efficiency of the pass depends on the nature of the transformation, its scale, its complexity and the transformation performance of the individuals nodes of the EMNN. EMNNs differentiate themselves from standard material structures by their ability to adapt to their environment and operator requirements. Training strategies for the material properties controlling the output of the inference pass are therefore central to their functioning. The degree of autonomous physical adaptability reflects how effectively the neural network-inspired structure physically approximates the desired continuous function without input from traditional computation.

\begin{figure}[h]
    \centering
    \includegraphics[width=0.9483\textwidth]{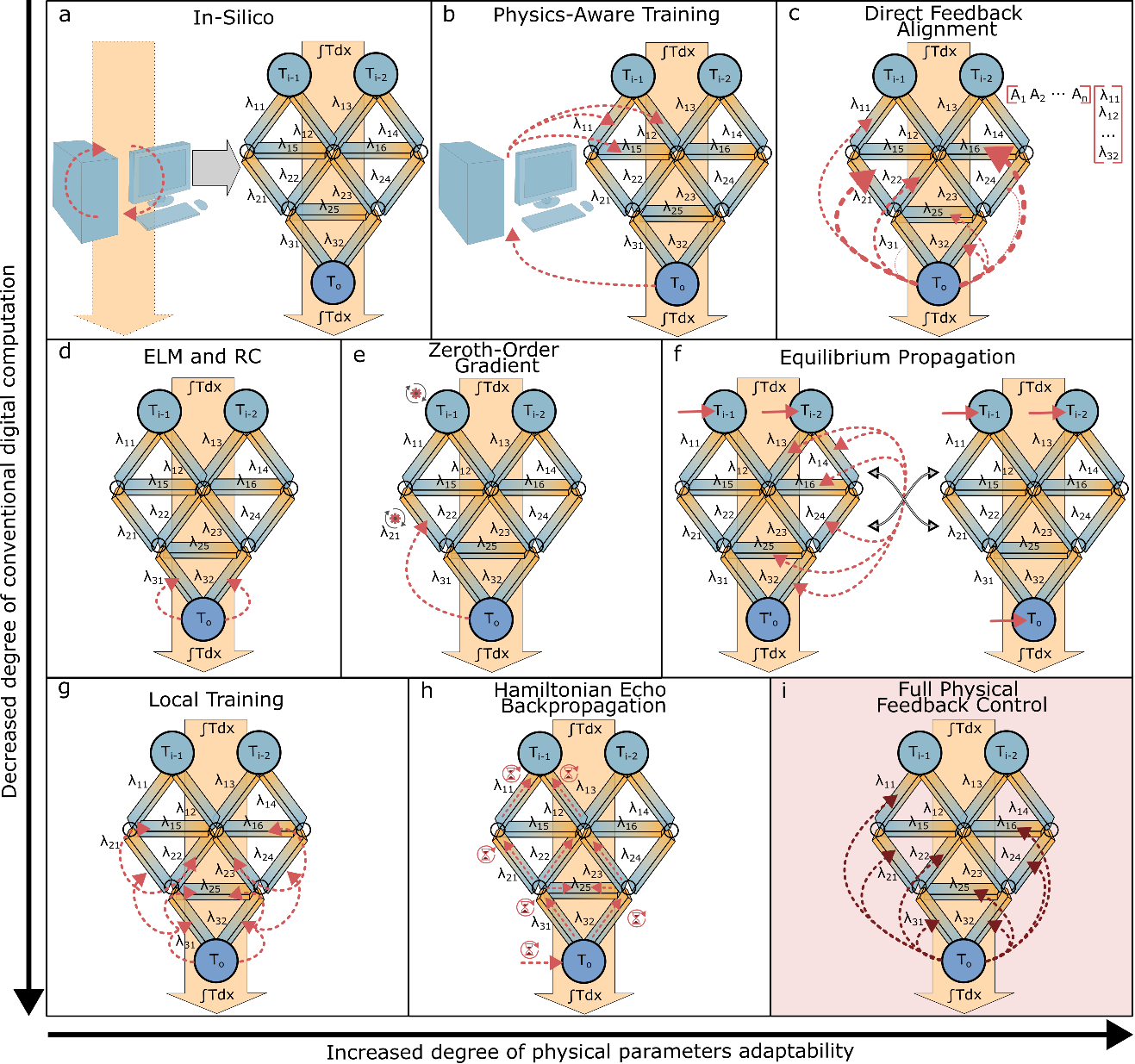}
    \vspace{-3mm}
    \caption{\textbf{Training strategies used and proposed for EMNNs.} The techniques the further on the right present increased adaptability. The techniques at the lower rows present more physical training with less intervention of traditional computing and human intervention (ELM: Extreme Learning Machine; RC: Reservoir Computing). Full Physical Feedback Control is not a training technique but rather the objective for the training methods for EMNN, fully controllable weight and independent from conventional training methods. The gear symbol highlights the optimisation of the weight of the materials involved from a perturbation of the input or the trainable weight. The hourglass symbol represent the time reversal update of the physical parameters.}
    \label{fig: 3}
\end{figure}

Non materials-based physical neural networks and metamaterials have different computation and adaptation strategies. Fig. \ref{fig: 3} ranks those different training strategies according to their degree of physical adaptability. The lowest degree of autonomous physical adaptability is \textbf{In-silico} training, where all computations are performed on GPUs before physical implementation \cite{Momeni2025TrainingNetworks} (Figure \ref{fig: 3}a). For instance, Li et al. use a type of in-silico training, the adjoint-based gradient descent backpropagation method. This method is used to train a stiffness-based mechanical neural network controlling output displacements \cite{Li2024TrainingBackpropagation}. However, the physical structure itself does not learn; its behaviour remains fully dependent on the pre-designed and manufactured properties of the mechanical metamaterial. Similar purely theoretical and computational models have been developed to design physical neural networks made of folded creased sheets, dynamic oscillating and topological mechanical neural networks \cite{Stern2020SupervisedSystem,deBos2026MultimodalProblem,Li2026TopologicalLearning}. The code used to design the folded creased sheet proved to be versatile and robust. This type of optimisation process used for classification also theoretically introduces additional material autonomous adaptability as seen in the physical neural network made from creased sheets \cite{Li2026TopologicalLearning}. In this model, the structures are initially produced untrained but later adapted to achieve the desired classification. Irreversible experimental learning has also been observed in similar polymeric-based local crease structures \cite{Arinze2023LearningBifurcation}. In this case, the polymer folding softens the crease and therefore supports the adaptation of the folding behaviour. The latter is an example of \textbf{Physics-Aware Training}  (Fig. \ref{fig: 3}b), which is more physical than in-silico training. The ultimate goal of Engineering Material Neural Networks is to physically and reversibly approximate the desired continuous function, entirely independent of external computation. However, developing in-silico-trained material neural networks provides a valuable design space to explore the feasibility of creating fully physical EMNNs.

Up to now, most of the Physical Neural Networks developed in open literature are based on physical mechanisms associated to light or sound, which can transmit information through air \cite{Momeni2025TrainingNetworks}. A large part of those physical neural network do not represent \textit{per se} Engineering Material Neural Networks, as the output is not dependent on the interaction between the information being transmitted and the intrinsic physical properties of a load bearing engineering material. However, optical neural networks, more advanced in development, offer valuable inspiration for EMNNs. These systems have been designed with strategies to make their structure incrementally physical and less digital \cite{Momeni2025TrainingNetworks}.

The challenge of creating an EMNN is the physical integration of these architectures with a training loop. Adjoint-based backpropagation serves as a technique bridging gradient-based backpropagation (subcategory of physics-aware training in Fig. \ref{fig: 3}b) and physics-based methods \cite{Momeni2025TrainingNetworks}. In a adjoint-based backpropagation-based configuration, the structure simulates the forward pass, but only solves an adjoint equation backward to update the weight of the code. The adjoint model illustrated in Fig. \ref{fig: 3}b presents a promising approach for introducing functional, physics-based learning in material neural networks. Another potential training strategy is presented in a theoretical study by Refinetti et al. and is named \textbf{Direct Feedback Alignment} training technique \cite{Refinetti2022AlignAlignment}. In this approach, individual weights are adjusted directly using injected random feedback matrices as shown in Fig. \ref{fig: 3}c, rather than propagating weight updates layer by layer. This method has the potential of limiting the backpropagation constrains from the EMNN, hence, facilitating their physical implementation. As shown in Fig. \ref{fig: 3}d, training EMNNs has the potential of being conducted by applying \textbf{Extreme Learning Machine} (ELM) \cite{Ermolaev2025LimitsMachine} and \textbf{Reservoir Computing} (RC) \cite{Tanaka2019RecentReview,Wright2022DeepBackpropagation} concepts, where not all weights are updated. For example, in certain architectures, it may be advantageous to train only the weights of the readout layer \cite{Hary2025PrinciplesFiber} with usually fixed reservoir weights. These methods are limited in adaptability, and will serve as an intermediary step toward effective learning throughout all layers via full physical feedback control training. Another intermediary solution is to design a link between inputs and outputs and optimise the layer weight for the target output. For instance, as described in Fig. \ref{fig: 3}e in \textbf{zeroth-order gradient}-estimation methods, trainable parameters are slightly perturbed and the resulting change in loss or output is measured to estimate an update direction. This differs from fully gradient-free search methods, which optimize without explicitly estimating gradients \cite{McCaughan2023MultiplexedBackpropagation}. To update the network’s weights, inputs or trainable weight are slightly perturbed, and the resulting output changes are measured and used to adjust the weights. \textbf{Equilibrium propagation} is a way of obtaining updates on the weights of the network by, for example, comparing the system in two different configurations. In general it consists in an energy-based local contrastive method to optimise weights. This typically requires two parallel instances of the system to simultaneously generate comparison points, eliminating the need for external memory \cite{McCaughan2023MultiplexedBackpropagation}. An example of this technique is described in Fig. \ref{fig: 3}f for EMNNs comparing to system one the output fixed and not the other. This method uses Hebbian adaptation and will be described as such in the coming paragraph. As shown in Fig. \ref{fig: 3}g a key principle for propagating information in a physical neural network is \textbf{physical local learning} updating neurons using only locally available information. This eliminates the need for full scale feedback control, reducing both equipment and physical space requirements for the material neural network structure \cite{Nkland2019TrainingSignals}. In such system, the loss is calculated locally with the actuation function of the specific layer involved and limits the physical equipment needed. This technique, already applied in acoustic, microwave and optical systems, holds significant potential for implementation in EMNNs. As described in Fig. \ref{fig: 3}h, \textbf{Hamiltonian Echo Backpropagation }uses a feedback system with an idealized energy model of the system to superimpose error onto the return signal, updating the structure \cite{Lopez-Pastor2023Self-LearningBackpropagation}. The arrows going though the grid structure in Fig. \ref{fig: 3}h indicate the update of the weights using the time reversal invariant dynamics of the Hamiltonian gradient update method. All the training methods previously highlighted present limitation regarding the objective of EMNNs to be as adaptable and independent from conventional computing systems. Therefore, Fig. \ref{fig: 3}i presents the \textbf{Full Physical Feedback Control} training for EMNN where each material is updated individually with physical mechanism. This last step is not a training method of a specific nature, it is the ultimate training target for the creation of EMNNs. In such structure all the weights are updated in a fully controllable manner by the operator and fully independently from conventional computing system.

\begin{figure}[h]
    \centering
    \includegraphics[width=0.83\textwidth]{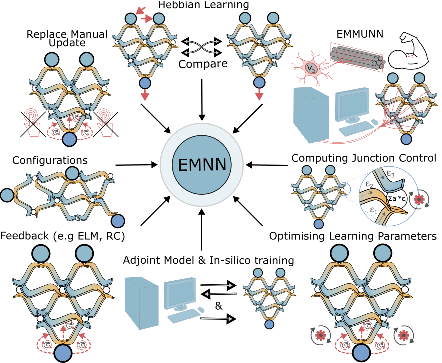}
    \vspace{-3mm}
    \caption{\textbf{Summary of research strategies highlighted in the present study to develop Engineering Material Neural Networks (EMNNs).} (ELM: Extreme Learning Machine ; RC: Reservoir Computing ; EMMUNN: Engineering Material Motor Unit Neural Network ; $\epsilon$: Strain). The gear symbol highlights a potential for updating optimisation for the weight of the material or the design of material connectors. The hand symbols represent the manual updates of parameters.}
    \label{fig: 4}
\end{figure}

To this extend, Fig. \ref{fig: 4} presents, alongside of the in-silico training, different potential directions of interest to explore for development of EMNNs. Fig. \ref{fig: 4} illustrates that, in addition to in-silico training, the adaptation of material properties can be made more physical by having in-silico analyse of their behaviour in parallel to the forward inference pass. For instance, Fig. \ref{fig: 3}b highlights a physics-aware training in which the forward pass is performed physically, but the learning and update of the weight is conducted in conventional computing. Such technique was proposed by Chen et al. to control the displacement of the tip of a beam by adjusting the stiffness of the trusses acting as a structural element for the beam itself \cite{Chen2025PhysicalEnvironments}. In this example, an \textit{Iterative Physical learning} strategy locally senses the position of the different trusses used as reinforcement in the beam and adjusts the stiffness of the beam on the difference between the local targeted position and the one measured. However, current training techniques for EMNNs and engineering metamaterials, beyond in-silico and physics-aware methods, are limited.

Mei et al. took an initial step toward autonomous material neural networks by developing a spring and buckled beams-based mechanical neural network. In this system, the controlled property is the output displacement, achieved through the bistable positioning of nodes \cite{Mei2024MechanicalNeurons}. Training calculations were performed in MATLAB using manually implemented gradient-based error correction. However, manually updating the network’s weights limits its physical adaptability. As illustrated in Fig. \ref{fig: 4}, a promising research direction for developing MNNs is to replace manual weight updates in existing physical neural networks with material-based mechanisms. Lee et al. proposed a mechanical neural network with the relation between input and output displacements controlled using adaptable  stiffnesses \cite{Lee2022MechanicalBehaviors}. The structure consisted of an assembly of tunable stiffness electronic beams. A gradient-based approach optimised beam stiffness to control output node displacement. Genetic algorithms and partial pattern search were also employed via \textit{in silico} training to determine the optimal stiffness configurations. While not a true Material Neural Network due to its hardware electronic components, the study provides valuable insights into in service stiffness control for such systems. This study was further extended through modelling, replacing the electronic tunable beams with bistable material structures \cite{Hopkins2023UsingLearn}. Modelling these structures enables adapting of training techniques originally developed for optical neural networks for mechanical systems. For example, Chen et al. propose a backpropagation-based training method for a lattice-based mechanical neural network, aimed at optimizing both static and dynamic behaviour \cite{Chen2024IntelligentBehaviors}. In Schaffland et al.’s educational mechanical neural network design, lever displacements serve as inputs, outputs and weights. Training is performed manually by the operator, loosely following a gradient-based approach \cite{Schaffland2023MechanicalEveryone}. Du et al. also employ experimental training in their mechanical neural network (though not explicitly labelled as such in their work). The system uses angular positions between nodes as inputs and produces the assembled node shape as the output \cite{Du2026MetamaterialsShape}. In this system, an adapted version of a physical local learning is applied, in which the angular position of a given node and its closest neighbour provide sufficient information to manually control the stiffness that counteracts the action of the motorised hinge. Local training is also applied in mechanical neural networks, where springs are manually adjusted at their endpoints to achieve a target position \cite{Altman2024ExperimentalNetworks}. Local learning can be also applied at the scale of the individual node. For example, Patil et al. developed nodes capable of autonomously adjusting the stiffness of an elastic band to match specific vibration patterns while ignoring others \cite{Patil2025ControlledComposites}. By combining these elements, the system can uncover new patterns of operation. However, many of the architectures proposed in the previous studies are not organised around a grid like structure, but rather along a line arrangement \cite{Du2026MetamaterialsShape}, or sets of irregular connections \cite{Altman2024ExperimentalNetworks}. As shown in Fig. \ref{fig: 4}, testing various node distributions is essential for developing physical neural networks, since spatial arrangement directly affects performance. Non grid structures have been proposed for resistor networks trained using an asynchronous learning rule, where nodes are updated individually rather than simultaneously\cite{Wycoff2022DesynchronousNetwork}. A theoretical framework to train locally different types of MNN with a required output is provided by Stern et al. \cite{Stern2021SupervisedMachines}. One proposed strategy draws inspiration from contrastive Hebbian adaptation (subcategory of Equilibrium Propagation presented in Fig. \ref{fig: 3}f), comparing two distinct states of the system \cite{Movellan1991ContrastiveModel}. One state consists in having the inputs fixed, while the other involves  both inputs and required outputs fixed. The network weights (e.g. stiffness, conductivity) are then locally updated using the difference between nodes outputs in both states. Hebbian adaptation and its derivatives have a great potential as training methods for material neural networks. Mechanical computing has also the potential to calculate weight updates autonomously for adaptation \cite{Yasuda2021MechanicalComputing,Hu2025SpatiallyRobotics}. Hu et al. propose a high computing density mechanical metamaterial as an assembly of logic gates for in service calculations \cite{Hu2025SpatiallyRobotics}. To achieve this training, the structure must be deformable into the desired output without sustaining damage.

In summary, material neural networks with training fully embedded in their materials properties do not yet exist. However, recent experimental and theoretical advances in related fields offer clear pathways for developing systems as those shown in Fig. \ref{fig: 3} and \ref{fig: 4}.

Before achieving full energy and computational autonomy, intermediate steps could include drawing inspiration from other neuron types (Fig. \ref{fig: 4}). Motor units are neurones actuating muscles in the body after being triggered by a centralised cortex system \cite{Heckman2012MotorUnit}. A material neural network could use a centralised control system: it would compare the target and actual output, then optimise node parameters via a neural network. Many metamaterials proposed today have the potential of becoming Engineering Material Motor Unit Neural Networks (EMMUNNs) by implementing physical training adaptability \cite{Luo2020DesignProperties,Poon2019Phase-ChangingMorphing}. The key advantage of these structures is the seamless integration of training and operation: as the system acquires data, it continuously fuels the EMNN’s real time functioning \cite{deKergariou2026PhysicalBiocomposites}.

\section{Materials}
\begin{figure}[h]
    \centering
    \includegraphics[width=0.983\textwidth]{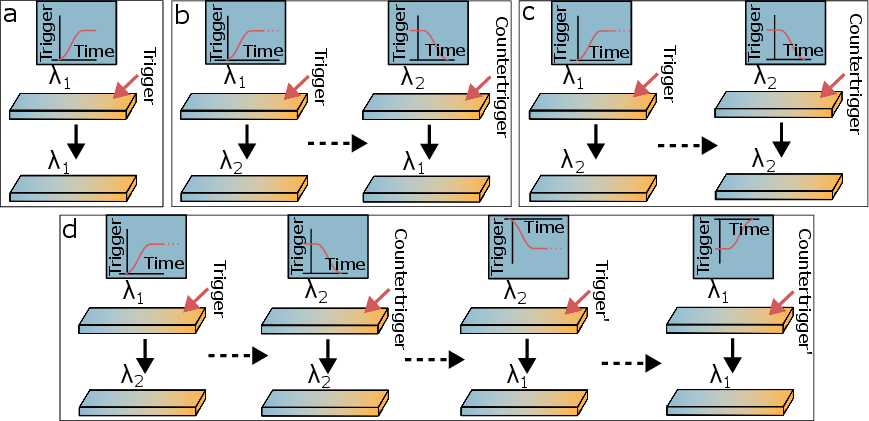}
    \vspace{-3mm}
    \caption{\textbf{Different degrees of adaptability and reversibility for materials in EMNNs.} ($\lambda$: material properties) \textbf{a,} Non adaptive material: the material properties do not change with the trigger. \textbf{b,} Non permanently adaptive material: the material properties change with the trigger, but when trigger is then removed (countertrigger), returning the material to its initial state. \textbf{c,} Permanently non reversible adaptive material: the material properties change with the trigger, but when the trigger is removed the properties stay with the new values. In this case, the trigger induces permanent changes in the material, leading to restrictions to the achievable range of properties of the material itself. \textbf{d,} Permanently reversible adaptive material: the material properties change with the trigger, but when the trigger is removed the properties stay with the new changed values. Properties of the material go back to the initial values if a different trigger is applied to the structure.}
    \label{fig: 5}
\end{figure}

The previous section describes systems either theoretical or with limited material physical adaptability out of traditional silicon-based computing. This section proposes practical, material-based solutions to enhance the adaptability of EMNNs. Walther proposes an energy and memory functional view to describe adaptable materials \cite{Walther2020Viewpoint:Roadmap}. However, to develop material neural networks, it is crucial to focus on how transitions between permanent states can approximate node behaviour. Fig \ref{fig: 5} illustrates the varying degrees of material adaptability required to develop EMNNs. Non adaptive materials in Fig. \ref{fig: 5}a cannot be used for EMNN, while the non-permanently adaptive materials of Fig. \ref{fig: 5}b do not intrinsically display permanent adaptability within the material, but can be used to generate EMMUNNs. Permanently non reversible adaptive materials of Fig. \ref{fig: 5}c can provide platforms to design EMNNs with limited permanent adaptability, as only certain changes of properties are possible. Permanently reversible adaptive materials, presented in Fig. \ref{fig: 5}d are ideal candidates to for EMNNs, as they provide full flexibility in changing the weights of the nodes. Such materials allow in theory to adapt the behaviour of the structure using linear or non linear materials. However, the feedback steps presented in Fig. \ref{fig: 2} will be harder to implement with non linear materials as the materials will be responsible for calculating and feed the updated weights back to the EMNN. Hence, non linearity in material limits the training methods available as it does for optics based neural Physical Neural Networks (PNNs) \cite{Momeni2025TrainingNetworks,Nakajima2022PhysicalHardware}. By adopting non linear material in EMNN, broken-isomorphism EMNN can be created if it does not reproduce operations of artificial neural networks. In such system the physical transformation of the material trains the system instead of having a set of operation mapped onto an artificial neural network.

\subsection{Examples of theoretical Engineering Material Neural Networks}

\begin{figure}[h]
    \centering
    \includegraphics[width=0.983\textwidth]{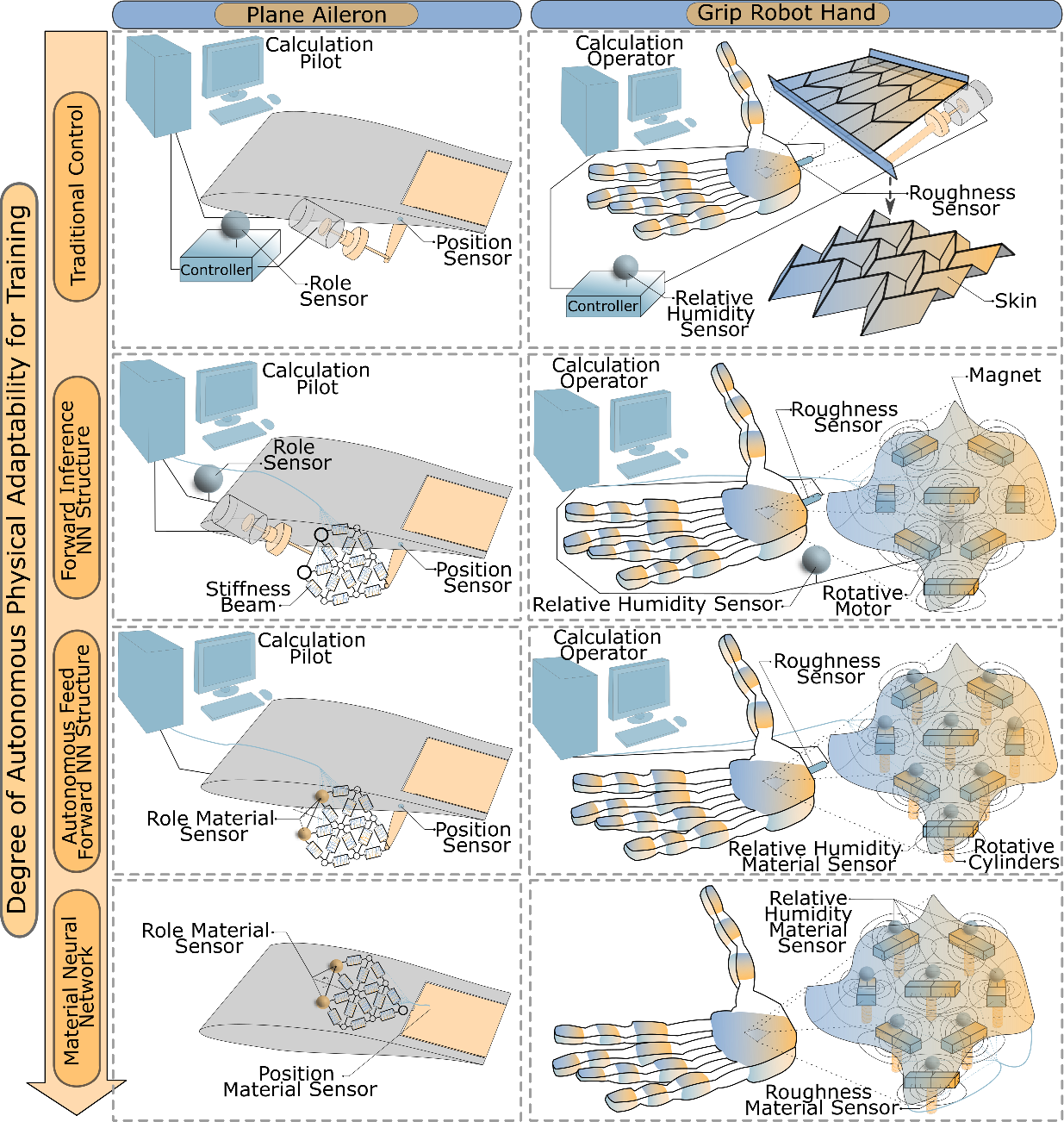}
    \vspace{-3mm}
    \caption{\textbf{Two examples of EMNNs with different degrees of material physical adaptability for training.} The \textit{Plane Aileron} example presents how EMNNs can be used to integrate controllable and adaptable stiffness beams to automatically control the position of an aileron on a plane's wing. The \textit{Grip Robot Hand} example introduces how an EMNN can be created out of an assembly of magnet to control the roughness of a soft skin to adapt the roughness of a robotic hand.}
    \label{fig: 7}
\end{figure}

This section illustrates theoretical EMNNs, bridging the gap between structure, concept and material properties. To clarify the requirements for developing EMNNs, we examine the thermal example in Fig. \ref{fig: 2}b. Here, the structure uses a polymer whose thermal conductivity can be dynamically adjusted through tensioning  \cite{Zhang2018NewFilms,Xu2026Strain-TunableDelocalization}. Xu et al. present, in particular, a polymer whose thermal conductivity can be reversibly tuned by adjusting the applied deformation \cite{Xu2026Strain-TunableDelocalization}. Using this material, the thermal conductivity ($\lambda_{ij}$), as shown in Fig. \ref{fig: 2}b, can be adapted in service. At this stage, the structure satisfies Rules $\#$1 and $\#$2 for MNNs and can be classified as a Feed Forward Neural Network Structure, as illustrated in Fig. \ref{fig: 7}. To qualify as a true engineering material neural network, the structure must fulfil user defined tasks and achieve adaptability and reversibility independently of digital computation. However, reaching this goal may involve intermediate development steps, inspired by the learning systems discussed earlier and the research directions outlined in Fig. \ref{fig: 4}. Initially, learning could be fully conducted using traditional computing. In this approach, electronic sensors detect the input and output temperatures as well as the strain in each material, while an electronic system dynamically stretches each material. A surrogate model is then implemented to predict and optimise the properties of all stretched materials under the desired conditions. To introduce additional material intelligence into the structure, Direct Feedback Alignment (DFA) can be employed to update the weights collectively. With this technique, as illustrated in Fig. \ref{fig: 7}d, a material is brought into direct contact with the output thermal structure of the system. Via Seebeck effect \cite{Atoyo2020EnhancedThermoelectrics}, a voltage is generated proportional to the temperature difference. As shown in Fig. \ref{fig: 7}g, this voltage is fed back into the structure’s materials via fixed random projections presented with the red box. By tuning the Seebeck properties, the voltage applied to each node’s weight can be precisely controlled. The resulting electrical current can be amplified and then directed to a piezoelectric material, which deforms to further modulate the material’s electrical conductivity. The training instrumentation is complex and must be miniaturised to minimise its impact on the material’s forward pass behaviour. To reduce interference from material update mechanisms, Extreme Learning Machine (ELM) can be employed \cite{Ermolaev2025LimitsMachine}, updating only the final weight layer.

Fig. \ref{fig: 7} illustrates two potential applications for Engineered Material Neural Networks, designed according to the previously defined structural framework. The application called \textit{Plane Aileron} is an extension of a concept proposed by Lee et al. \cite{Lee2022MechanicalBehaviors}, in which stiffness-based neural networks (called mechanical neural networks) could be implemented on blades \cite{Lee2022MechanicalBehaviors}. The second example in Fig. \ref{fig: 7}, called \textit{Grip Robot Hand}, describes the possibility to generate wrinkles on the surface of humanoid robots to mimic human skin, with the wrinkles improving the handling of wet objects in immersed conditions \cite{Kareklas2013Water-inducedObjects}. 

The generation of Engineering Material Neural Networks would require the completion of four stages of development (Fig. \ref{fig: 7}). The first stage is the \textit{Traditional Control} one, where both training and implementation rely entirely on conventional computing. This approach does not require embedded material intelligence and does not align with any of the three rules of the EMNN concept outlined above. In the blade example, the aileron’s orientation is determined by an engine, which processes the pilot’s input, roll sensor data, and the aileron’s current position. Here, the material itself has no intelligence as all computations are external. The skin wrinkling robot follows a similar principle: a computer measures skin roughness and ambient humidity, while the engine adjusts the skin’s texture based on sensor data and operator requirements such as the object to be grasped. No embedded matter intelligence is also needed in this case. 

The second step towards the creation of EMNN discusses the discretisation of the controlling system (\textit{Forward Inference NN Structure} in Fig. \ref{fig: 7}). At this stage of the automation, rules \#1 and \#2 previously highlighted for Mechanical Neural Network are respected. In the aileron example, variable stiffness materials are assembled to convert the engine’s input into precise control of the aileron’s position. The engine’s input may remain constant or vary, but the EMNN’s stiffness properties are dynamically adjusted to achieve the desired output. In this case, the system needs initial training before being able to be used as controlling system. Similarly, in the robotic hand example, the skin embedded with metallic particles is controlled using discrete magnets interconnected through their magnetic fields. Thus, by modulating the magnetic field intensity of the magnets (e.g. for instance via through humidity) various skin configurations can be achieved. Controlling the orientation of a single magnet enables large scale skin configurations, as adjusting one magnet’s alignment influences the entire magnetic network. Training is needed to map inputs (e.g., humidity, object to grasp) to desired outputs (e.g., skin roughness) via control parameters (e.g., magnetic field intensity). In both systems, the structure then achieves the desired outputs in a forward pass implementation. The autonomous forward inference pass NN configuration of Fig. \ref{fig: 7} can represent the structure of a Material Motor Unit Neural Network (MMUNN) in the case of \textit{Calculation Pilot} (See Fig \ref{fig: 7}) and \textit{Calculation Operator} (See Fig \ref{fig: 7}) using centralised neural network to obtain an optimised version of each node for the configuration and conditions at stake.

The third step consists in adding autonomy to a forward inference NN-like structure (\textit{Autonomous Forward Inference NN Structure} in Fig. \ref{fig: 7}). For the aircraft aileron, input no longer comes from an engine but from the material’s interaction with its environment. In the blade example, the material-based roll sensor must harness acceleration to generate its own input energy. A mass undergoing motion can generate energy with a roll acceleration, similar to the pendulum used in robotics \cite{Patil2023Self-learningCircuits}. A similar principle is applied in anti-roll tanks on boats, where energy generated by roll motion is harnessed to dampen lateral movement \cite{Kapsenberg2023AShips}. Lee et al. specifically proposed to use the pressure generated by the air on the wing to actuate the mechanical neural network they designed \cite{Lee2022MechanicalBehaviors}. Regardless of its source, the generated energy must be calibrated to serve as input for the discretised stiffness structure described earlier. For the robotic hand, inspired by heliotropic sunflowers and bat wings, 4D printing technology has enabled the development of rotary, programmable cylinders \cite{Soleimanzadeh2025RotaryMandrel} capable of rotative motion due to environmental changes. The magnet’s motion in the \textit{Grid Robot Hand} application presented in Fig \ref{fig: 7}  will thus be directly driven by these systems in response to humidity levels. At this stage, in both systems, material properties (analogous to neural network weights) are still computed based on system and environmental responses, then adjusted via an electronic system. Therefore, the adaptability of the behaviour of the material is still computed and implemented with conventional electronics. However, the structure now begins to adhere to \textit{Rule \#3} of MNNs, as input control becomes independent of traditional methods.

Finally, the last step yields a fully realised \textit{Material Neural Network}, fulfilling all the previously defined rules. To achieve this for the blade example in Fig. \ref{fig: 7}, a Position Material Sensor (e.g., a lever system) is proposed to track the aileron’s position. A method is required to compute the positional error signal, the difference between the target and actual positions, and adaptively update the layers of the material neural network. One potential effective approach involves pre-training the layers before finalising the structure as a forward inference Engineering Material Neural Network (EMNN). For example, as detailed later in this study a temporary training apparatus can potentially be attached to the discretised stiffness elements of the MNN to calibrate their properties before removal for in service operation. In this training step, optimisation of the meta parameters can be conducted to find, for instance, the ideal sensitivity to update the weights of the neural network. In this case the structure would not be an in service learning EMNN but rather a deployed device with a trained inference structure. The robotic hands, however, can become an in-service learning EMNN. To achieve this, the EMNN must use a responsive material that senses its wrinkling shape, compares it to the required deformation (e.g., gripping surface wrinkles), and adjusts the magnetic field accordingly to achieve precise control.

\subsection{Materials for Engineering Material Neural Networks}
\begin{figure}[h]
    \centering
    \includegraphics[width=0.9983\textwidth]{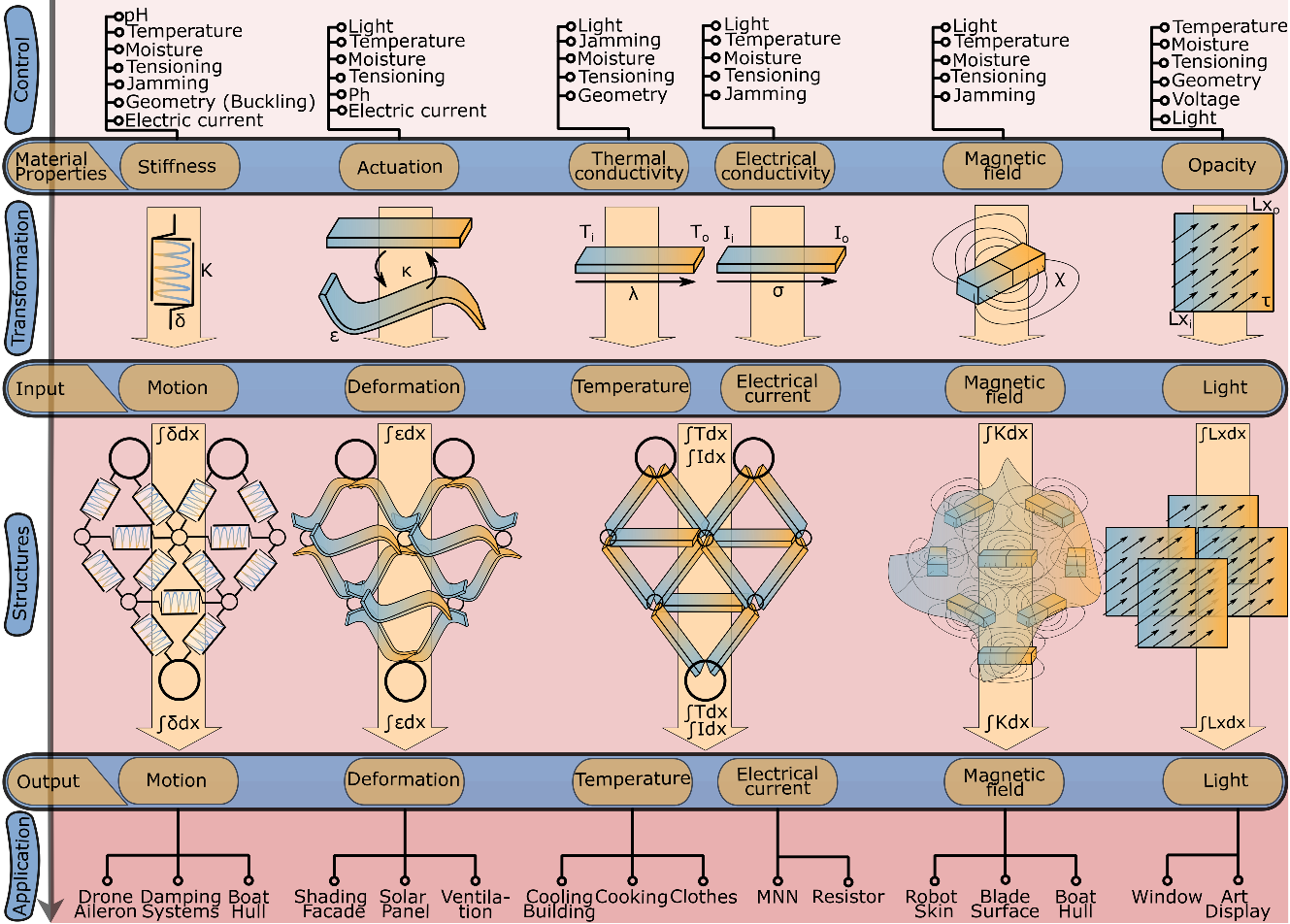}
    \vspace{-5mm}
    \caption{\textbf{Subcategories of Engineering Material Neural Networks.} The row labelled as \textit{Materials Properties} presents the parameters of the materials to be modified in the training phase of the neural network-like structure ($\delta$: displacement ;K: stiffness; $\kappa$: curvature; $\epsilon$: displacement; T: Temperature; $\lambda$: thermal conductivity; $\sigma$: electrical conductivity; I:Electrical current intensity; $\chi$: Magnetic field intensity; $\tau$: opacity; Lx: light intensity). Different ways to control these properties are described in the \textit{Control} section of the figure. The rows labelled as \textit{Input} and \textit{Output} show the parameters controlled via the NN-like structure. Potential applications for these adaptive structures are also provided.}
    \label{fig: 6}
\end{figure}

Building on the theoretical EMNN example from the previous section, Fig. \ref{fig: 6} introduces various subcategories of Material Neural Network concepts, each tailored to specific material systems. As in Fig. \ref{fig: 2}a, these designs feature repeated material patterns arranged in a grid, linking inputs to outputs through distinct material properties. The figure also highlights potential applications for these adaptive structures.

To evaluate the feasibility of the concepts presented in Fig. \ref{fig: 6}, it is essential to identify and analyse the key material properties necessary for their realisation. Past studies have shown the possibility to control stiffness using pH \cite{Lulevich2004EffectMicrocapsules,Deng2025ReprogrammableEncoding}, temperature \cite{Owuor2018HighTransparency}, moisture \cite{deKergariou2022TheComposites,Deng2025ReprogrammableEncoding}, tensioning \cite{Shah2016DamageLoading,Vehoff2007MechanicalRelaxation}, jamming \cite{Coulais2014ShearJamming,Liu2025RateMetamaterials}, electric fields \cite{Wei2011VibrationBeams}, vibration \cite{Wang2021Bio-inspiredCrosslinking} and geometry/buckling \cite{Janbaz2019Ultra-programmableMechanisms,Kuppens2021MonolithicMachines,Yan2024Self-deployableProperties}. The actuation methods employed to modify the deformation of materials follow similar principles as those used in adaptive material systems. For instance, pH can be used \cite{Xiao2016Light-Effect}, as well as temperature \cite{Chen2025HybridWriting}, humidity \cite{deKergariou2022DesignHygromorphs,Deng2025ReprogrammableEncoding}, electrical currents\cite{Delbart2024DevelopmentSimulation} and mechanical constraints \cite{deKergariou2025Hygromnemics:PreConstraining} to control actuation authority. The thermal conductivity of materials used in the MNN presented in Fig. \ref{fig: 2} can be adapted with the surrounding humidity \cite{Emperhoff2024OnHydrogen}, tensioning \cite{Li2010StrainNanostructures}, light \cite{Shin2019Light-triggeredPolymers}, geometry \cite{Luo2022OrientationPrinting} and jamming \cite{Xue2023SegregatedParticles}. Modifications of the electrical conductivity can be obtained using light \cite{Yigzaw2025EffectPEDOT:PSS}, temperature \cite{Sanchez-Trujillo2023TemperatureFilms}, geometry/orientation \cite{Yao2023SimulationMicro/Nanocomposites}, humidity \cite{Khurram2025InfluenceFilms}, loading/tensioning \cite{Ciampaglia2025ExperimentalMonitoring} and jamming \cite{Vitelli2010HeatSolids}. Magnetic fields can be affected by jamming configurations of magnet-based composites \cite{Aktas2025JammingComposites}. The magnetisation of a material can also be changed with temperature and humidity \cite{Ohkoshi2004Humidity-inducedAssembly}. but also with mechanical loading via magnetostriction \cite{Lee1955MagnetostrictionEffects} and with light via photomagnetism \cite{Pejakovic2000PhotoinducedMagnet}. The right column in Fig. \ref{fig: 6} shows how to control opacity with temperature \cite{Ye2025ThermoresponsiveBehaviors}, humidity \cite{Xiang2021Humidity-DrivenWindow}, tensioning \cite{Jiang2019DynamicConditions}, light \cite{Zhao2022EntirelyStorage} and voltage \cite{Gu2022EmergingDisplays}. Most of the materials discussed here are conventional bulk engineering materials, offering limited scope for precise property tuning. However, by integrating industrial-grade electronic chips with tailored properties, the control precision of material neural networks can be significantly enhanced \cite{Luo2024ResponseMetamaterial}. An alternative approach is to replace single bulk materials per node with material assemblies, enabling multiple configurations, and thus diverse outputs from varied inputs. Shimohara et al. produced a compliant mechanism which achieves a binary stiffness by combining materials \cite{Shimohara2023CompliantFreedom}. However, this approach lacks compactness, potentially leading to spatial constraints in MNNs. As shown in Fig. \ref{fig: 6}, materials in MNNs are assembled using connectors (black circles). A critical challenge in their design is ensuring that these connectors introduce minimal disruption to the intrinsic properties of the materials. For instance, in thermal MNNs (Fig. \ref{fig: 2}), the connectors must possess thermal conductivity several orders of magnitude greater than that of the nodes, while in stiffness-based MNNs, the connectors must be multiple orders of magnitude stiffer than the nodes themselves.

The material becomes non-adaptive if its properties are pre-designed during manufacturing \cite{Xue2023SegregatedParticles,Luo2022OrientationPrinting,Yao2023SimulationMicro/Nanocomposites}. Table \ref{tab: 1} categorises the materials shown in Fig. \ref{fig: 5}, including those discussed in this study when non-adaptive. It also underscores the limited availability of permanently reversible adaptive materials (PRAMs). The literature reveals that most permanently reversible adaptive materials (PRAMs) rely on jamming and self healing mechanisms for property control. Tensioning, particularly when inducing plastic deformation, serves as a promising method for achieving permanently non-reversible adaptive materials (PNRAMs). For PNRAMs, composite materials are essential, as their distinct phases enable the integration of multiple required functions within a single material system. An example of PNRAM composite materials have been provided by Orrego et al., who have proposed a composite material made of piezoelectric PVDF that generates electrical current by calcifying ions on its surface when loaded, thus adding stiffness to the structure \cite{Orrego2020BioinspiredProperties}. In this case, the PVDF serves as trigger for the stiffening effect, while the mineral deposited on the PVDF serves as the mechanical reinforcement. Piezoelectric material can have their microsctructure precisely control for instance via additive manufacturing to introduce control in the relationship between mechanical and electrical signal \cite{Sharma2025PiezoelectricManufacturing}.

One of the main groups of materials with the potential to create PRAMs are microstructured \cite{Bhaskar2009MicrostructuredFibers,Sakshi2025RoboticActuation} and micro-architected materials \cite{Peng2023MachineMaterials}. They are designed with small scale structures able to generate specific properties at large scale. Different patterns such as micropyramids, micropillars and microdomes are used to achieve this scaling up properties \cite{Sakshi2025RoboticActuation}. Precise control of the internal geometry of these materials is crucial to enable their intelligent functions within EMNNs \cite{Gao2023RationalProspects}. For example, multiscale material architectures can be leveraged to govern both forward pass and material updating behaviours of the structure via the multiple material they consist of \cite{Bhaskar2009MicrostructuredFibers,Peng2023MachineMaterials,Gao2023RationalProspects}. Different materials have the potential to be used for the different functions of the EMNN. By engineering the interactions between these scales, it becomes possible to dynamically regulate the updates of individual nodes, enhancing the adaptive performance of the EMNNs. An example of microstructure material benefiting from multiple scales has been produced as a bistable magnetic ceramic reinforced resin composite \cite{LeFerrand2019FilteredTransduction}. The material created in this study answers few of the requirements for the example applications presented in Fig. \ref{fig: 6} \textit{Grip Robot Hands} is the humidity trigger is replaced by mechanical one. When subjected to mechanical load from its environment, the composite responds by buckling, which alters both its magnetic properties and electrical conductivity. In this study a PRAM (see Fig. \ref{eq: 4}) is created as a mechanical trigger change the magnetic properties. Furthermore, the countertrigger does not alterate the magnetic properties obtained. Finally, a trigger of the same nature than the initial one leads to a return to the initial magnetic properties \cite{LeFerrand2019FilteredTransduction}. The main limitation shared by this material and most documented in the literature is the small number of stable states, here limited to just two \cite{LeFerrand2019FilteredTransduction}. In this system, the shift in magnetic properties can drive changes in the EMNN’s forward inference (e.g., altering surface roughness, as in Fig. \ref{fig: 6}), while the change in electrical conductivity can serve as a positional sensor (e.g., analogous to the relative humidity sensor in Fig. \ref{fig: 6}). The functionality of a microstructured material can be tailored by adding chemical elements and groups that trigger desired functions. For instance, Li et al. propose a multi responsive sandwich actuator in which the microstructure is made out of C, Si, Ga and Fe elements, which provide potential for thermal, magnetic actuation and electrical conductivity \cite{Li2025ClimbingSurfaces}. The resulting bending behaviour can be used in the \textit{Actuation} example in Fig. \ref{fig: 6}. Bending necessarily induces stretching and compression of the material. In a non-symmetrical sandwich structure like the one presented by Li et al. \cite{Li2025ClimbingSurfaces}, the change of relative dimensions between different layers has the potential to become a sensing parameter of the deformation. Liquid crystal elastomers also have possible use in PRAM for EMNN applications. For example, Lahikainen et al. present an azobenzene-based liquid crystal polymer actuator whose actuation behaviour can be dynamically reconfigured in service by exposing the material to different electromagnetic wavelengths\cite{Lahikainen2018ReconfigurableEffects}.

Engineering Living Materials are also a potential source of permanently non-reversible and reversible adaptative materials \cite{Nettersheim2024EngineeredMaterials}. Neural-like engineered living materials can be understood as distributed systems in which embedded cells function analogously to neurons, sensing environmental inputs and processing information locally \cite{Nguyen2018EngineeredMaterials}. In natural systems such as bone, osteocytes act as mechanosensors and communicate through interconnected networks by releasing signalling molecules, coordinating system-level responses without centralized control \cite{Bonewald2011TheOsteocyte}. Similarly, in engineered living materials, genetically engineered cells can be programmed with responsive gene circuits that translate stimuli into specific outputs, effectively acting as biological activation functions. Dranseike et al. emphasize the possibility to create actuating PNRAM engineered living materials via addition of bacteria inside hydrogel \cite{Dranseike2025DualMaterials}. Shining light to the encapsulated bacteria stiffens the structure. Through cell–cell communication and feedback, these systems enable collective decision making and adaptive behaviour, where repeated stimuli can reshape material properties. This combination of distributed sensing, signalling, and adaptive remodelling positions engineered living materials as a material level analogue of neural networks, capable of processing information and evolving their responses over time. For example, chloroplast can be used to locally stiffen hydrogel with light, thus creating a PNRAM material \cite{Yu2021Photosynthesis-assistedStructures}. Kharbedia et al. also prepared a polyacrylamide matrix with reversible stiffness behaviour via the addition of Mg$^{2+}$ to FtsZ living polymers inside. therefore creating a PRAM \cite{Kharbedia2026ExtensilePolymers}. By leveraging engineered genetic circuits, Nguyen et al. demonstrated a transition from reactive substances to history-dependent materials that effectively encode a memory of past environmental stimuli \cite{Nguyen2018EngineeredMaterials}. This biological adaptation loop enables the material to modulate its own production and morphology based on previous exposure, mirroring the "training" phase of a neural network. Within the EMNN framework, such cellular mechanisms serve as the foundation for autonomous, metabolic computation, where the material’s future state is a direct function of its experiential history. Most of the examples given in the engineered living materials section here are based on soft hydrogel materials. Further research is needed to scale up this material property control method to engineering materials and structures. One way the engineered living materials paradigm can be expanded consists in using synthetic genetic switches and optogenetic control to embed sequential logic and spatio-temporal memory directly into the material \cite{Schmachtenberg2025Self-assemblingCircuits}. Such circuit can encodes logic into the material providing potential for creating PRAM. For instance, when coupled with engineered structural material systems \cite{Mohammadi2019BiomimeticReinforcements}, these systems could evolve into physical reservoir computers where the material’s structural architecture and metabolic history converge to autonomously process complex environmental data. Mohammadi et al. controlled the properties of the silk during production but further development has the potential of controlling the properties in service \cite{Mohammadi2019BiomimeticReinforcements}. Mycelial matter also has potential to become PNRAM and PRAM, for instance, via damage control of properties. Damage in material can create discontinuity in the structure modifying any of the controlled material properties described in Fig. \ref{fig: 6}. By embedding fungal hyphae within 3D printed hydrogels, the mycelium functions as a decentralized, regenerative logic gate whilst providing structural integrity. This biohybrid interface moves beyond simple repair, utilizing nutrient-driven growth to autonomously re-optimize its structural integrity and "remember" its mechanical set-points after trauma \cite{Gantenbein2022Three-dimensionalMaterials}. Hence, mycelium-based material create a framework for engineered living materials PRAM with the first trigger being the displacement creating internal damage and the second trigger being the displacement to join the material parts together and the time to build the mycelium network again. Adding functional particle to the mycelium based material allow to enhance its multi functional properties, such as ions transfer \cite{Cheng2026Self-grownDevices} bioelectric signal and mechanical properties \cite{Schyck2026ShapingMaterials}. Such property control enhancement will be necessary to turn this technology to engineering materials. Beyond sensing local material scale stresses, fungal electrical signals have also been used in the control of robotic systems, presenting a pathway to integrate living matter decision making with more conventional mechatronic systems \cite{Mishra2024SensorimotorMycelia}. Across these engineered living materials works, some key themes emerge; from genetic memory which can store past inputs, growth-driven reinforcement (through precipitation, densification or alignment), stress responsive gene expression (bioelectricity leading to protein expression), reorganisation and collective cellular behaviours (emergent learning). These themes can enable a Hebbian-like adaptability process where the material’s response evolves with experience. As a result, the material can process inputs, store history, and adapt its future behaviour, functioning as a distributed, trainable biological network.

Finally, the combination of biological materials and bio-inspired designs represent key pathways for developing permanently reversible adaptive materials. For example, Matsuda et al. developed a hydrogel inspired by muscle adaptation, which strengthens in response to mechanical loading \cite{Matsuda2019MechanoresponsiveTraining}. This material has the potential to be used as PNRAM material. Researchers have also engineered bone-inspired materials that stiffen under dynamic vibrational loading \cite{Wang2021Bio-inspiredCrosslinking}. Another mechanical trigger with the potential for PRAM material are two ways shape memory behaviour \cite{Kim2023ShapeEffects,deKergariou2025Hygromnemics:PreConstraining}. Biological systems, such as cephalopod camouflage, also offer inspiration for materials whose properties can be dynamically controlled through multiple triggers. For instance, Xu et al. developed a non permanent adaptive material whose optical properties can be modulated by both mechanical tensioning and electrical stimulation \cite{Xu2020StretchableSystems}.

\begin{table}[h!]
\small
\centering
\caption{\label{tab: 1} \textbf{Degree of adaptability and reversibility of the materials referenced in the present study.} NPAM (Non Permanently Adaptative Material), PNRAM (Permanently Non Reversible Adaptative Material) and PRAM (Permanently Reversible Adaptative Material)}
\begin{tabular}{p{3.383cm} p{2.983cm} p{3.083cm} p{3.283cm} }
\hline
\begin{small} Properties\end{small}  &  \begin{small} NPAM\end{small} & \begin{small} PNRAM \end{small} & \begin{small} PRAM \end{small}  \\ \hline
\begin{small} Stiffness \end{small}  & \begin{small}  \cite{Lulevich2004EffectMicrocapsules,Owuor2018HighTransparency,deKergariou2022TheComposites,Deng2025ReprogrammableEncoding} \end{small} & \begin{small} \cite{Shah2016DamageLoading,Vehoff2007MechanicalRelaxation,Nettersheim2024EngineeredMaterials,Wang2021Bio-inspiredCrosslinking,Orrego2020BioinspiredProperties,Liu2025RateMetamaterials,Yu2021Photosynthesis-assistedStructures} \end{small}  & \begin{small} \cite{Coulais2014ShearJamming,Janbaz2019Ultra-programmableMechanisms,Kuppens2021MonolithicMachines,Yan2024Self-deployableProperties,Kharbedia2026ExtensilePolymers,Schyck2026ShapingMaterials,Gantenbein2022Three-dimensionalMaterials} \end{small}    \\
\begin{small} Actuation\end{small}  & \begin{small} \cite{Xiao2016Light-Effect,Chen2025HybridWriting,Deng2025ReprogrammableEncoding,Li2025ClimbingSurfaces,Sharma2025PiezoelectricManufacturing} \end{small} & \begin{small} \cite{deKergariou2025Hygromnemics:PreConstraining} \end{small}  & \begin{small} \cite{LeFerrand2019FilteredTransduction,Lahikainen2018ReconfigurableEffects,Kim2023ShapeEffects} \end{small}   \\
\begin{small} Thermal Conductivity\end{small}  & \begin{small}  \cite{Emperhoff2024OnHydrogen,Li2010StrainNanostructures,Shin2019Light-triggeredPolymers} \end{small} & \begin{small}  \end{small}  & \begin{small}  \end{small}    \\
\begin{small} Electrical Conductivity\end{small}  & \begin{small} \cite{Yigzaw2025EffectPEDOT:PSS,Sanchez-Trujillo2023TemperatureFilms,Khurram2025InfluenceFilms,Ciampaglia2025ExperimentalMonitoring,Ye2025ThermoresponsiveBehaviors} \end{small} & \begin{small}  \end{small}  & \begin{small} \cite{Ciampaglia2025ExperimentalMonitoring} \end{small}    \\
\begin{small} Magnetic Field\end{small}  & \begin{small} \cite{Ohkoshi2004Humidity-inducedAssembly,Lee1955MagnetostrictionEffects}  \end{small} & \begin{small} \cite{Pejakovic2000PhotoinducedMagnet} \end{small}  & \begin{small} \cite{Aktas2025JammingComposites} \end{small}    \\
\begin{small} Light\end{small}  & \begin{small} \cite{Ye2025ThermoresponsiveBehaviors,Xiang2021Humidity-DrivenWindow,Jiang2019DynamicConditions,Gu2022EmergingDisplays,Xu2020StretchableSystems} \end{small} & \begin{small}  \end{small}  & \begin{small} \cite{Zhao2022EntirelyStorage} \end{small}   \\\hline
\end{tabular}
\end{table}

As described previously, the adaptation of the material is key to creation of EMNNs.However, it is equally critical to either precisely compute the output or adapt it during training to match the desired result. In many cases, this step will include calculations at material level. The possibility to calculate for a material has been demonstrated in recent years \cite{Louvet2025ReprogrammableModes,Fang2016PatternCompute}. However, these materials are designed for a specific task, such as pattern recognition\cite{Fang2016PatternCompute} or multiplication\cite{Louvet2025ReprogrammableModes}. To apply similar computational techniques, the material must be tailored to each EMNN whether for recognising output patterns or distributing updates across nodes, for example. To advance Engineered Material Neural Networks, materials science must explore three critical directions:

\begin{itemize}
    \item The development of Permanently Reversible Adaptative Materials.
    \item The compaction of the training adaptative properties (e.g. in the micro structure; with living materials; with biological materials) feedback loop in the structure of Permanently Reversible Adaptative Materials used in EMMNs.
    \item The adapt materials capable of computing to control the forward pass and feedback control stages of EMNNs. Create new computing materials specific to EMNNs.
\end{itemize}

\section{Conclusion}
A groundbreaking shift in materials science involves embedding intelligence directly into the structure and properties of materials, intelligence can be defined as the ability of a material to approximate functions taking environmental parameters and operator objectives as inputs and producing adaptive behavioural responses as outputs. A network of interconnected nodes can be considered as a basis to approximate continuous functions. Thus, assembling engineering materials as interconnected node networks enables them to adapt to their environment and generate the desired output. This principle designing such adaptive assemblies is defined here as the first rule for constructing Material Neural Networks (MNNs) from engineering materials. The first rule requires the ability to individually tailor the properties of each material. The second rule demands that the network’s output adapt based on operator intent. Finally, the third rule stipulates that the Engineered Material Neural Network (EMNN) should be designed to optimise energy efficiency, output accuracy, independence from traditional digital computations and speed of convergence.

By theoretically defining the structure of Engineered Material Neural Networks (EMNNs) and reviewing training techniques for non-material-based physical neural networks, we have identified key research directions for advancing the development of Material Neural Networks. For example, we propose a less physically constrained EMNN by introducing Material Motor Unit Neural Networks (MMUNNs), inspired by the motor unit neurons in biological systems. These neurons translate signals from the central nervous system into muscle responses. Hence, the feedback control in MMUNNs is made from in-service control of non permanently adaptative material. The development of adjoint models and \textit{in silico} training aims to extend this concept to novel materials with limited material-based feedback control, bridging the gap between biological adaptability and engineered systems. The development of feedback loops, computationally capable material junctions, and optimized adaptation parameters aims to expand the information transmission capabilities of existing Engineered Material Neural Networks (EMNNs). Exploring the learning potential of EMNNs involves testing various configurations, applying Hebbian adaptability principles, and replacing manual updates with physically or digitally implemented update rules with an established relationship to the defined loss or objective. Finally, materials were classified into distinct categories—based on their adaptability and the reversibility of their property changes to determine their suitability for EMNN applications. Composite materials, engineered living materials, microstructured materials, microarchitected materials, and biological materials were identified as holding significant potential for developing EMNNs.
In the materials science context, the primary conclusion of this work is to emphasise the urgent need for computing capable, reversible and permanently adaptive materials essential building blocks for advancing EMNN technology for a sustainable society.

\section{Acknowledgements}
C.d.K., A.W.P and F.S. acknowledge the support of the ERC-2020-AdG101020715 NEUROMETA project. The corresponding author, C.d.K.,also thanks the EPSRC Doctoral Prize Fellowship for supporting this work (Grant No. EP/W524414/1).

\section{Conflict of interest}
No author declare any conflict of interest.

\end{document}